\documentclass[12pt,a4paper,final]{iopart}
\usepackage[latin9]{inputenc}
\usepackage[latin9]{inputenc}
\usepackage{color}
\usepackage{amsmath}
\usepackage{amssymb}
\usepackage{graphicx}
\usepackage{esint}

\usepackage[unicode=true,pdfusetitle,
 bookmarks=true,bookmarksnumbered=false,bookmarksopen=false,
 breaklinks=true,pdfborder={0 0 1},backref=false,colorlinks=true]
 {hyperref}
\hypersetup{
 linkcolor=red,urlcolor=blue,citecolor=red,anchorcolor=blue}
\usepackage{breakurl}

\makeatletter
\@ifundefined{textcolor}{}
{%
 \definecolor{BLACK}{gray}{0}
 \definecolor{WHITE}{gray}{1}
 \definecolor{RED}{rgb}{1,0,0}
 \definecolor{GREEN}{rgb}{0,1,0}
 \definecolor{BLUE}{rgb}{0,0,1}
 \definecolor{CYAN}{cmyk}{1,0,0,0}
 \definecolor{MAGENTA}{cmyk}{0,1,0,0}
 \definecolor{YELLOW}{cmyk}{0,0,1,0}
}

\usepackage{times}
\usepackage{psfrag}
\usepackage{epsf}
\usepackage{epsfig}\usepackage{latexsym}\usepackage{amsfonts}
\usepackage{latexsym}
 \usepackage{ifpdf}

\begin{document}

\title{Quasi one-dimensional Bose-Einstein condensate in gravito-optical surface trap}

\author{Javed Akram$^{1,2}$}
\address{$^1$Institute f\"{u}r Theoretische Physik, Freie Universit\"{a}t Berlin, Arnimallee
14, 14195 Berlin, Germany}

\address{$^2$Department of Physics, COMSATS, Institute of Information Technology
Islamabad, Pakistan}

\ead{javedakram@daad-alumni.de}

\author{Benjamin Girodias$^{3,4}$}

\address{$^3$Department of Physics and Astronomy, Pomona College, California,
USA}
\address{$^4$Department of Physics, University of Michigan, Michigan, USA}

\ead{bgiro@umich.edu}

\author{Axel Pelster$^{5}$}
\address{$^5$Fachbereich Physik und Forschungszentrum OPTIMAS, Technische Universit\"{a}t
Kaiserslautern, Germany}

\ead{axel.pelster@physik.uni-kl.de}

\begin{abstract}
We study both static and dynamic properties of a weakly interacting
Bose-Einstein condensate (BEC) in a quasi one-dimensional gravito-optical
surface trap, where the downward pull of gravity is compensated by
the exponentially decaying potential of an evanescent wave. First,
we work out approximate solutions of the Gross-Pitaevskii equation
for both a small number of atoms using a Gaussian ansatz and a larger
number of atoms using the Thomas-Fermi limit. Then we confirm the
accuracy of these analytical solutions by comparing them to numerical
results. From there, we numerically analyze how the BEC cloud expands
non-ballistically, when the confining evanescent laser beam is shut
off, showing agreement between our theoretical and previous experimental
results. Furthermore, we analyze how the BEC cloud expands non-ballistically
due to gravity after switching off the evanescent laser field in the
presence of a hard-wall mirror which we model by using a blue-detuned
far-off-resonant sheet of light. There we find that the BEC shows
significant self-interference patterns for a large number of atoms,
whereas for a small number of atoms, a revival of the BEC wave packet
with few matter-wave interference patterns is observed. 

\end{abstract}

\pacs{67.85.Hj, 03.75.Kk}
\vspace{2pc}
\noindent{\it Keywords}: Bose-Einstein condensate, matter-wave interference, quasi one-dimensional, gravito-optical
surface trap

\section{Introduction}

Bose-Einstein condensation (BEC) is impossible in a one- or a two-dimensional
homogeneous system \cite{PhysRevLett.17.1133,PhysRev.158.383}, but
does occur in atomic traps because the confining potential modifies
the density of states \cite{PhysRevLett.87.130402,PhysRevLett.87.080403,Petrov04}.
Experimentally, a highly elongated quasi-1D regime can be reached
by tightly confining the atoms in the radial direction, which can
be effectively achieved by letting the radial frequency be much larger
than the axial frequency \cite{PhysRevLett.91.250402,PhysRevLett.91.010406,PhysRevLett.92.190401,PhysRevLett.95.190406,PhysRevLett.95.260403,HofLesSch07,Eckart08,Pethick02}.
However, when the radial length scales become as the order of the
atomic interaction length, the one-dimensional system can only be
described within the Tonks-Girardeau or within the super Tonks-Girardeau
regime \cite{PhysRevLett.81.938,PhysRevLett.85.3745,PhysRevLett.91.163201},
which is experimentally reachable near a confinement-induced resonance
\cite{Paredes04,Kinoshita04,Haller09}.

The interaction of lower dimensional ultra-cold atoms with surfaces
has attracted much attention in the past few years as their enhanced
quantum and thermal fluctuations have turned out to play an important
role for various technological applications \cite{PhysRevLett.95.190403,PhysRevA.80.021602,Vetsch12,PhysRevA.90.023607}.
Also under such circumstances, the influence of gravity must be taken
into account. For many years, atomic mirrors were constructed in the
presence of a gravitational field by using repulsive evanescent waves,
which reflect both atomic beams and cold atom clouds \cite{COOK1982258,Balykin1987151,Liston95}.
The trapping of atoms in a gravitational cavity, consisting of a single
horizontal concave mirror placed in a gravitational field, is discussed
in detail in Refs. \cite{Wallis92,Wallis96}. The inherent losses
of atoms in a gravitational cavity can be reduced by using a higher
detuning between the evanescent wave and the atomic resonance frequency
in a gravitational trap \cite{PhysRevLett.71.3083}. In 1996, Marzlin
and Audretsch studied the trapping of three-level atoms in a gravito-optical
trap by using the trampolining technique without the trampoline \cite{PhysRevA.53.4352}.
Another approach proposed by Saif et al., uses a spatially periodic
modulated atomic mirror to yield either a localization \cite{PhysRevA.58.4779}
or a coherent acceleration \cite{Akram08} of material
wave packets depending on the chosen initial conditions and the respective
system parameters. In 2002, Nesvizhevsky et al. reported the evidence
of gravitational quantum bound states of neutrons in a gravitational
cavity \cite{Nesvizhevsky02}. Not only is a good confinement geometry
necessary for trapping and observing the dynamics of atoms, but an
experiment also needs an efficient loading scheme loading scheme.
The experimental group of Rudi Grimm from Innsbruck demonstrated both
the loading of $^{133}$Cs atoms \cite{PhysRevLett.79.2225,Hammes00}
and the subsequent creation of a BEC in a quasi-2D gravito-optical
surface trap (GOST) \cite{Domokos01,PhysRevLett.92.173003}. More
recently, Colombe et al. studied the scheme for loading a $^{87}$Rb
BEC into a quasi-2D evanescent light trap and observed the diffraction
of a BEC in the time domain \cite{Colombe03,PhysRevA.72.061601}.
Later Perrin et at. studied the diffuse reflection of a BEC from a
rough evanescent wave mirror \cite{Perrin06}.

\begin{figure}
\begin{centering}
\includegraphics[width=3.2in,height=2.3in]{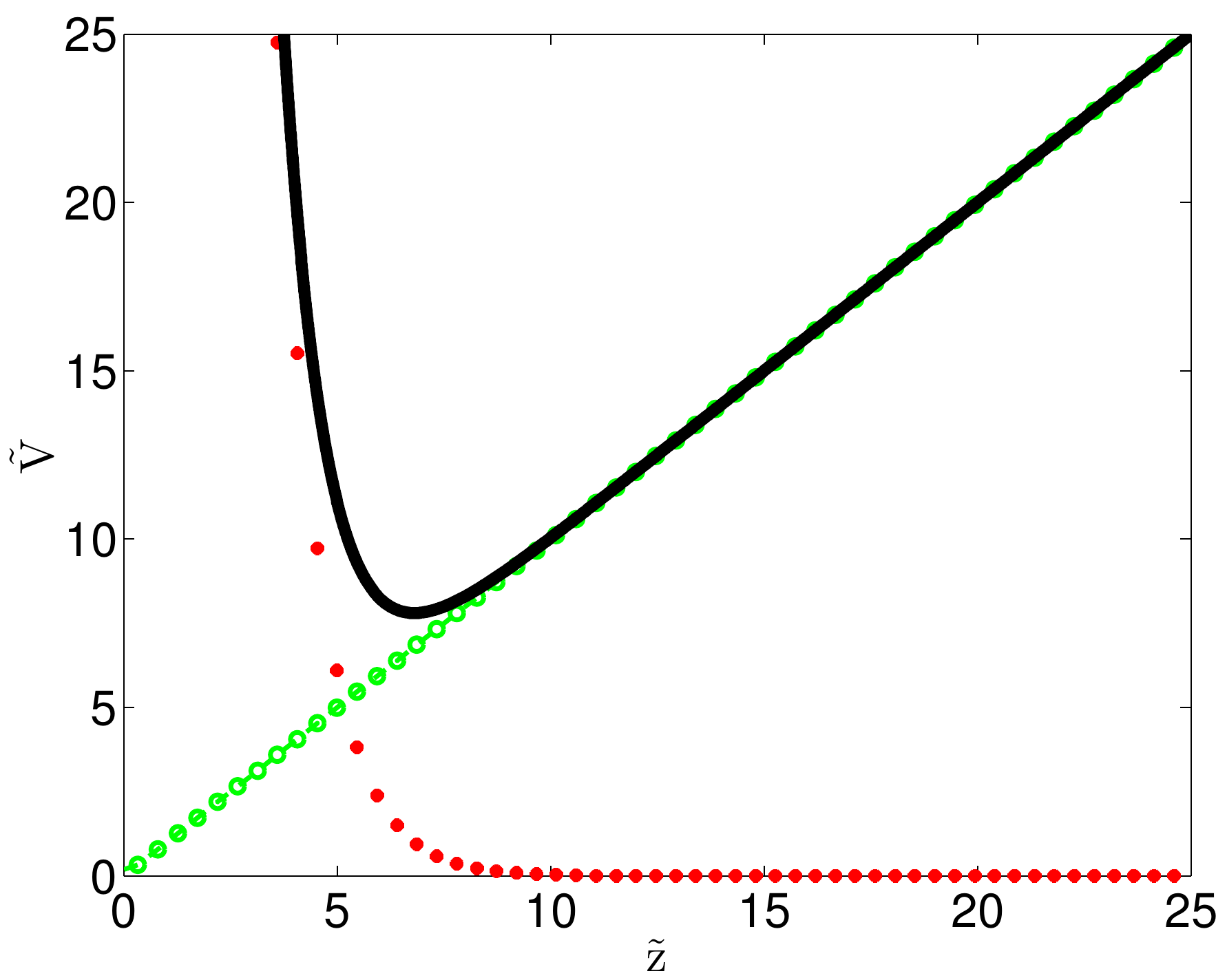} 
\par\end{centering}

\protect\protect\protect\protect\protect\protect\protect\caption{(Color online) Anharmonic GOST potential (solid line) from Eq. (\ref{Eq2})
in dimensionless units, which are explained at the end of the Sec.
\ref{Model}. It consists of a superposition of an exponentially decaying
optical potential (red circles) due to an evanescent light field above
a mirror and a linear gravitational potential (dashed-green circles).
When the atoms are cold enough, they stay in the vicinity of the potential
minimum. }

\label{Poten} 
\end{figure}

Motivated by the crucial relevance of gravito-optical surface traps
in atomic waveguides \cite{Gattobigio10,PhysRevA.84.042712,PhysRevA.86.062713}
and atomic chips \cite{PhysRevLett.94.090405,PhysRevA.83.043406,Petrovic13,Jian14},
in this paper we study the special case of a quasi one-dimensional
Bose-Einstein condensate which is trapped orthogonal to the prism
surface along the vertical axis. In our proposed model the downward
pull of gravity is compensated by an exponentially decaying evanescent
wave (EW), which can be thought of as a mirror as it repels the atoms
upward against gravity as shown in Fig. \ref{Poten}. In order to
deal with the hard-wall boundary condition, we apply the mirror solution
analogy to the BEC context, and obtain analytical results, which agree
with those from numerically solving the underlying one-dimensional
Gross-Pitaevskii equation (1DGPE). Later on, as an interesting application,
we compare our numerical simulation results for a time-of-flight dynamics
with the Innsbruck experiment for a quasi-2D BEC in a GOST \cite{PhysRevLett.92.173003}.
Surprisingly, our proposed quasi one-dimensional model agrees even
quantitatively with the Innsbruck experiment. Although this Innsbruck
experiment uses a 2D pancake shaped BEC, when performing the time-of-flight
expansion vertically the transversely confining beam is kept constant,
so our quasi-1D model for a BEC should apply in this case. 

The paper is organized as follows, the underlying fundamental model
for such a quasi-1D BEC is reviewed in Sec. \ref{Model}. Furthermore,
we provide estimates for experimentally realistic parameters, which
we use in our quantitative analysis. Afterwards, we work out approximate
solutions for the 1DGPE wave function in the ground state of the system.
To this end, Sec. \ref{Ansatz} performs a modified Gaussian variational
ansatz for weak interactions, which corresponds to a small number
of atoms. For a larger number of atoms, the interaction strength becomes
so strong that the Thomas-Fermi solution turns out to be valid, as
described in Sec. \ref{TF}. Then, in Sec. \ref{Num}, we outline
our numerical methods and compare them to the previous analytical
solutions. In Sec. \ref{TOF}, we deal with the time-of-flight expansion
of the BEC when the EW is removed, showing quantitative agreement
with previous experimental results. In Sec. \ref{Dynamics} we discuss
further dynamical properties of the BEC in a GOST after switching
off the evanescent laser field in the presence of the hard-wall mirror.
Lastly, we summarize our findings and end with brief concluding remarks.

\section{Model}

\label{Model} For our 1D model of the BEC in a GOST, we assume that
we have a dilute Bose gas and that the radial frequency is much larger
than the axial frequency, i.e. the BEC is cigar-shaped. With this
assumption, we arrive at the 1DGPE \cite{Kamchatnov04,Gonzalez08}
\begin{equation}
i\hbar\frac{\partial}{\partial t}\psi(z,t)=\left\{ -\frac{\hbar^{2}}{2m_{\text{B}}}\frac{\partial^{2}}{\partial z^{2}}
+V(z)+\text{G}_{\text{B}}\lvert\psi(z,t)\rvert^{2}\right\} \psi(z,t).\label{TDGP}
\end{equation}
On the right-hand side of the equation, the first term represents
the kinetic energy of the atoms with mass $m_{\text{B}}$, while the
last term describes the two-particle interaction, where its strength
$\text{G}_{\text{B}}=2N_{\text{B}}a_{\text{B}}\hbar\omega_{r}$ is
related to the s-wave scattering length $a_{\text{B}}$, and the particle
number $N_{\text{B}}$, whereas $\omega_{r}$ denotes the radial trapping
frequency. The anharmonic potential energy $V(z)$ in Eq. (\ref{TDGP})
is produced by both gravity and the exponentially decaying evanescent
wave as shown in Fig.~\ref{Poten} \cite{PhysRev.158.383}: 
\begin{equation}
V(z)=V_{0}e^{-\kappa z}+m_{\text{B}}\text{g}z.\label{Eq2}
\end{equation}
Here, $\text{g}$ is the gravitational acceleration and the constant
$V_{0}=\Gamma\lambda_{0}^{3}I_{0}/(8\pi^{2}c\delta_{3})$ denotes
the strength of the evanescent field, where $\Gamma$ is the natural
linewidth of $^{133}$Cs atoms, $\lambda_{0}=852~{\rm {nm}}$ is the
wavelength of the optical transition, $I_{0}$ stands for the peak
intensity of the EW, and $\delta_{3}$ corresponds to the detuning
frequency of the hyperfine sub-level $F=3$ of the $^{133}$Cs atom
\cite{PhysRevLett.79.2225,Hammes00,PhysRevLett.92.173003}. Furthermore,
${1/\kappa=\Lambda/2=\lambda/4\pi\sqrt{n^{2}\sin^{2}\theta-1}}$ represents
the decay length, where $\lambda$ is the wavelength of the EW, $n$
stands for the refractive index of the medium and $\theta$ is the
angle of incidence. The potential Eq. (\ref{Eq2}) has a minimum at
$z_{0}^{{\rm {min}}}=(1/\kappa)\ln(V_{0}\kappa/m_{\text{B}}\text{g})$
with the axial frequency $\omega_{z}=\sqrt{\text{g}\kappa}$. Note
that this potential yields a hard-wall condition with $V(z\leq0)=\infty$,
because the atoms cannot penetrate the prism, as it is a macroscopic
object.


In order to have a concrete set-up in mind for our analysis, we adapt
parameter values from the GOST experiments \cite{PhysRevLett.79.2225,PhysRevLett.92.173003}.
For the EW, we consider the inverse decay length to be $\kappa=2/\Lambda=1.43\times10^{6}~{\rm {m^{-1}}}$,
i.e. $\Lambda\approx1.4\,{\rm {\mu m}}$. Additionally, we assume
an axial frequency of ${\omega_{z}\approx2\pi\times600~{\rm {Hz}}}$.
For our atoms in the $F=3$ state, the strength of the EW is given
by $V_{0}\approx100\times k_{B}~{\rm {\mu K}}$, where $k_{B}$ is
the Boltzmann constant. This potential value is within an order of
magnitude of the Innsbruck experiments \cite{PhysRevLett.79.2225,PhysRevLett.92.173003}.
In view of the quasi-1D model, we must satisfy the condition $\omega_{\text{z}}\ll\omega_{\text{r}}$,
so we assume ${\omega_{r}=2\pi\times3~{\rm {kHz}}}$, which corresponds
to the radial oscillator length $l_{\text{r}}=0.892~{\rm {\mu m}}$.
The experiment uses a magnetic field for Feshbach resonance, such
that the s-wave scattering length amounts to $a_{\text{B}}=440~{\rm {a}_{0}}$
with the Bohr radius $a_{0}$. As both $\omega_{\text{z}}\ll\omega_{\text{r}}$
and $l_{\text{r}}>a_{\text{B}}$ are fulfilled, we have, indeed, a
quasi one-dimensional setting.

When the atoms are close to the dielectric surface of the prism, we
would have to add an additional potential contribution due to the
van-der-Waals interaction. Here we have to distinguish two special
cases depending on the distance of the BEC from the surface. The regime,
where the BEC is close to the dielectric surface, i.e. $z\ll\lambda/2\pi$,
is called the Lennard-Jones regime \cite{Jones32,PhysRevA.53.1862}.
In the case where the BEC is far away from the surface, i.e. $z\gg\lambda/2\pi$,
the regime is called Casimir-Polder regime. In the latter case the
Casimir-Polder potential can be described within a two-level approximation
of the Cesium atoms according to $V_{{\rm {CP}}}=-C_{4}/z^{4}$, where
the Casimir-Polder coefficient is $C_{4}=1.78\times10^{-55}{\rm {Jm^{4}}}$
\cite{PhysRev.73.360,PhysRevLett.77.1464,PhysRevLett.98.063201}.
As the BEC does not penetrate very far into the repulsive EW, it is
hardly influenced by the van-der-Waals potential, which follows from
the above values of the GOST experiment. Indeed, for the wavelength
$\lambda=839~{\rm {nm}}$ \cite{PhysRevLett.92.173003} and the distance
being estimated by the minimal distance of the BEC from the surface
$z_{0}^{{\rm {min}}}=4.761~{\rm {\mu m}}$, we are in the the Casimir-Polder
regime. Thus, the value of the Casimir-Polder potential is of the
order $V_{{\rm {CP}}}=24.9\times{\rm {k_{B}}~{\rm {pK}}}$, which
is negligible in comparison with the EW potential $V_{0}\approx100\times k_{B}~{\rm {\mu K}}$.

\begin{figure}[h]
\begin{centering}
\includegraphics[width=3.2in,height=2.3in]{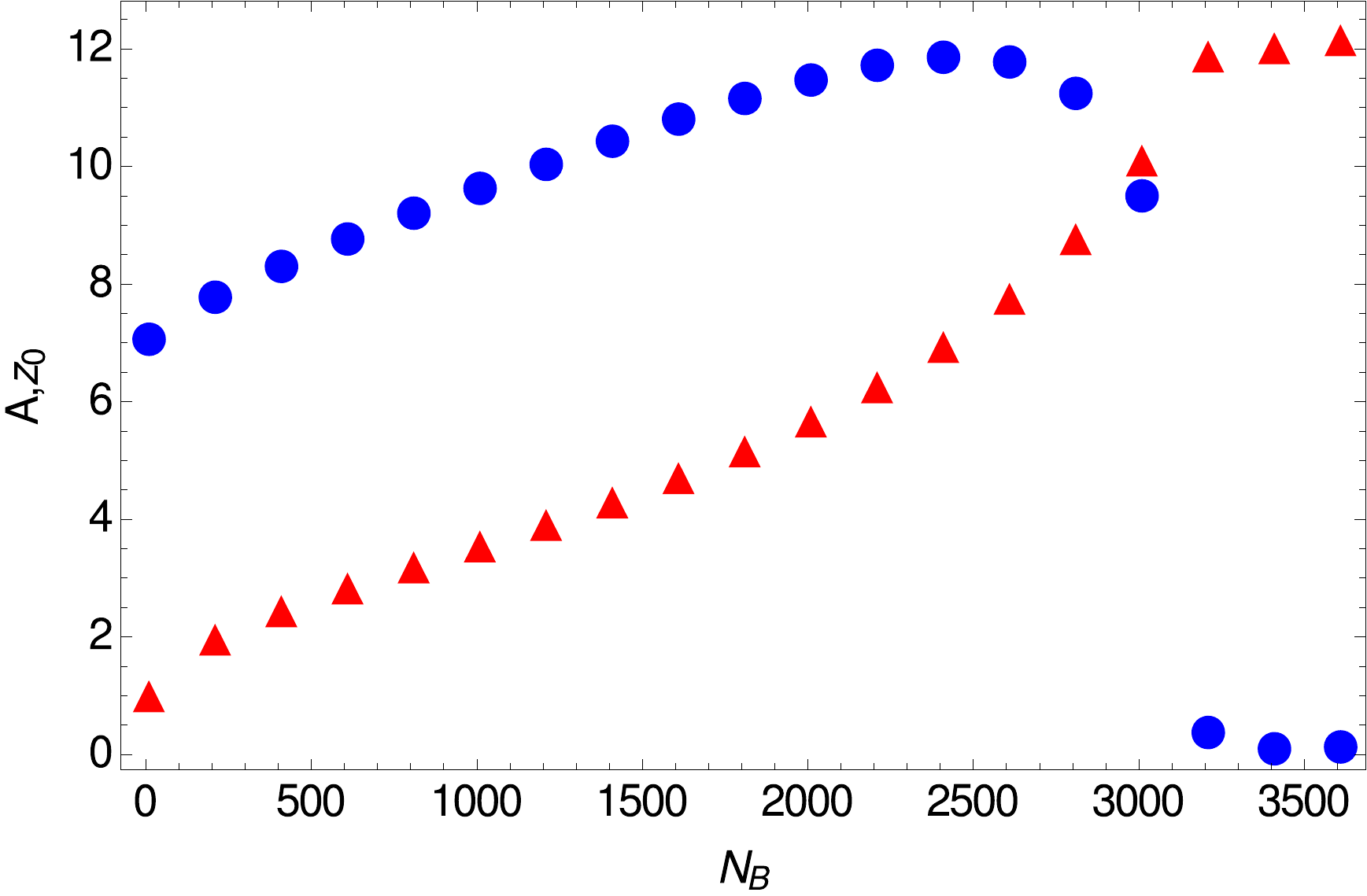} 
\par\end{centering}

\protect\protect\protect\protect\protect\protect\protect\caption{(Color online) Width $A$ (triangles) and mean position $z_{0}$ (circles)
as a function of the number of atoms $N_{\text{B}}$. Note that the
mean position gives meaningless values for $N_{\text{B}}>3000$. Thus,
the variational ansatz is only successful for quite a small number
of particles.}

\label{AnsatzParam} 
\end{figure}

In view of the forthcoming discussion we use dimensionless parameters
as follows. First we introduce the dimensionless spatial coordinate
$\tilde{z}=\kappa z$. Further, we multiply all terms in Eq. (\ref{TDGP})
by $\kappa/(m_{\text{B}}\text{g})$, yielding the dimensionless GPE,
\begin{align}
i\frac{\partial}{\partial\tilde{t}}\tilde{\psi}(\tilde{z},\tilde{t})= & \left\{ -\frac{k}{2}\frac{\partial^{2}}{\partial\tilde{z}^{2}}
+\tilde{z}+\tilde{V_{0}}e^{-\tilde{z}} +\tilde{\text{G}}_{\text{B}}\lvert\tilde{\psi}(\tilde{z},\tilde{t})\rvert^{2}\right\} \tilde{\psi}(\tilde{z},\tilde{t}),\label{nondim}
\end{align}
where the dimensionless kinetic energy constant reads $k=(\hbar^{2}\kappa^{3})/(\text{g}m_{\text{B}}^{2})$,
the dimensionless time $\tilde{t}=t(m_{\text{B}}\text{g})/(\hbar\kappa)$
and the two-particle dimensionless interaction strength $\tilde{\text{G}}_{\text{B}}=2N_{\text{B}}\tilde{\omega}_{r}\tilde{a}_{\text{B}}$
with $\tilde{a}_{\text{B}}=a_{\text{B}}\kappa$ being a dimensionless
s-wave scattering length. Additionally, we measure energies in units
of the gravitational energy $m_{\text{B}}\text{g}/\kappa$ and get
$\tilde{\omega}_{z}=\hbar\kappa\omega_{z}/\text{g}m_{\text{B}}$ as
a dimensionless frequency, and $\tilde{V}_{0}=\kappa V_{0}/\text{g}m_{\text{B}}$
as a dimensionless strength of the evanescent field.

According to these chosen parameters, the dimensionless quantities
have the following values. The dimensionless strength of the EW is
$\tilde{V}_{0}=905.7$, the dimensionless kinetic energy amounts to
$\tilde{k}=0.066$, the dimensionless s-wave scattering length is
given by $\tilde{a}_{\text{B}}=0.033$, the dimensionless radial frequency
yields $\tilde{\omega}_{r}=1.303$, and, finally, the resulting dimensionless
two-particle interaction strength is $\tilde{\text{G}}_{\text{B}}=0.086N_{\text{B}}$.
From here on, we will drop the tildes for simplicity.

\section{Variational solution}

\begin{figure}[h]
\begin{centering}
\includegraphics[width=3.2in,height=2.3in]{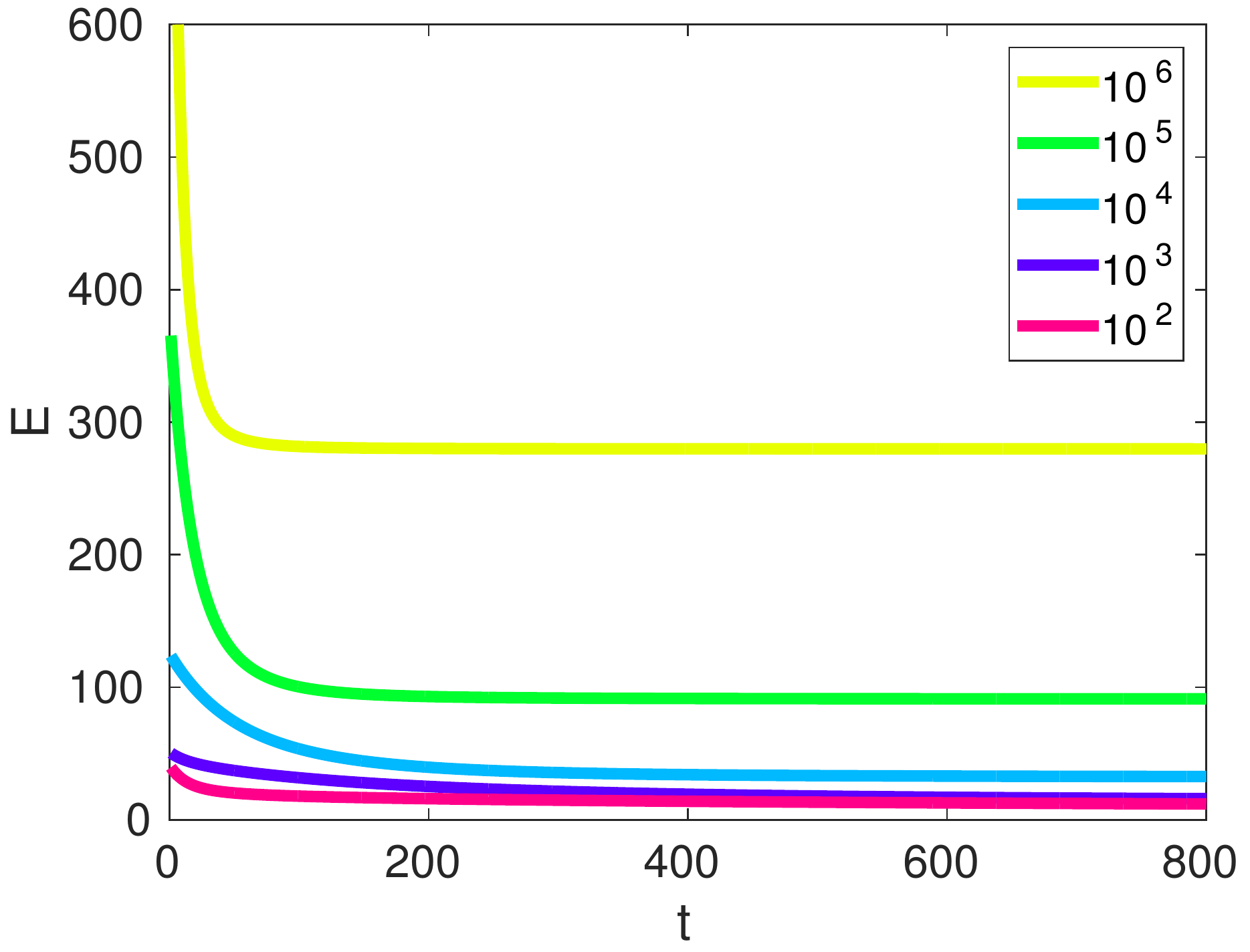} 
\par\end{centering}

\protect\protect\protect\protect\protect\protect\protect\caption{(Color online) Energy of the BEC as a function of imaginary time for
a decreasing number of particles from the top to the bottom. }

\label{converge} 
\end{figure}

\label{Ansatz} For the number of particles $N_{\text{B}}<3000$,
the effective interaction strength is quite small. In this limit the
BEC has only a reduced extension, so the anharmonic confinement $V(z)$
approximately corresponds to a harmonic potential well around its
minimum $z_{0}^{{\rm {min}}}\approx6.809$. Therefore, it is reasonable
to propose a Gaussian-like ansatz for the wave function in the static
case. In order to meet the hard-wall condition, however, we modify
the Gaussian function such that it has the form of a so-called mirror
solution \cite{Robinett06,Belloni14}, 
\begin{equation}
\psi(z)\propto\exp\left[\frac{-(z-z_{0})^{2}}{2A^{2}}\right]-\exp\left[\frac{-(z+z_{0})^{2}}{2A^{2}}\right],\label{eq:4}
\end{equation}
where $z_{0}$ is the mean position and $A$ represents the width.
In this way it is guaranteed that the wave function meets the hard-wall
boundary condition $\psi(0)=0$. In order to find the variational
parameters $z_{0}$ and $A$, we minimize the energy of this ansatz
following the idea of Perez et al. \cite{PhysRevLett.77.5320,PhysRevA.56.1424}.
In order to simplify the expression for the corresponding energy,
we introduce the parameter ${\gamma=z_{0}/A}$ and normalize the wave
function (\ref{eq:4}) to obtain 
\begin{equation}
\psi(z)=\frac{\exp\left(-\frac{z^{2}}{2A^{2}}\right)\text{Sinh}\left(\frac{\gamma z}{A}\right)}{\sqrt{\frac{A}{4}\sqrt{\pi}[\exp\left(\gamma^{2}\right)-1]}}.\label{eq:5}
\end{equation}
From this ansatz, we obtain the Gross-Pitaevskii energy

\begin{eqnarray}
E =\frac{V_{0}\{e^{\frac{1}{4}(A-2\gamma)^{2}}\left[\text{Erfc}\left(\frac{A}{2}-\gamma\right)+e^{2A\gamma}\text{Erfc}\left(\frac{A}{2}+\gamma\right)\right]
-2e^{\frac{A^{2}}{4}}\text{Erfc}\left(\frac{A}{2}\right)\}}{2\left(e^{\gamma^{2}}-1\right)}
\notag\label{eq:6}\\
 +\frac{2Ae^{\gamma^{2}}\gamma\text{Erf}(\gamma)}{2\left(e^{\gamma^{2}}-1\right)}+\frac{\left(2\gamma^{2}
 +e^{\gamma^{2}}-1\right)k}{4A^{2}\left(e^{\gamma^{2}}-1\right)}+\frac{\left[2\left(e^{\frac{\gamma^{2}}{2}}+1\right)^{-2}
 +1\right]\text{G}_{\text{B}}}{2\sqrt{2\pi}A}.\:\:\:\label{eq:61}
\end{eqnarray}
where $\text{Erf}(y)=\frac{2}{\sqrt{\pi}}\int_{0}^{y}e^{-x^{2}}dx=1-\text{Erfc}\left(y\right)$
denotes the error function. Although, this expression cannot be minimized
analytically, we can use numerical techniques to extremize it with
respect to the parameters $\gamma$ and $A$ based on the values of
$k$, $\text{G}_{\text{B}}$, and $V_{0}$ given in Sec. \ref{Model}.
We can see from {Fig.~\ref{AnsatzParam}} that our variational
approach turns out to be valid only for quite a small number of atoms.
Indeed, the extremization process fails when the condensate has more
than around $3000$ atoms, as then the mean position becomes zero
as shown in {Fig.~\ref{AnsatzParam}}. Note that BEC experiments
with such small particle numbers are possible, see for instance Refs.
\cite{Gerton2000a,Colombe07}.

\begin{figure}[h]
\begin{centering}
\includegraphics[width=5.5in,height=2.3in]{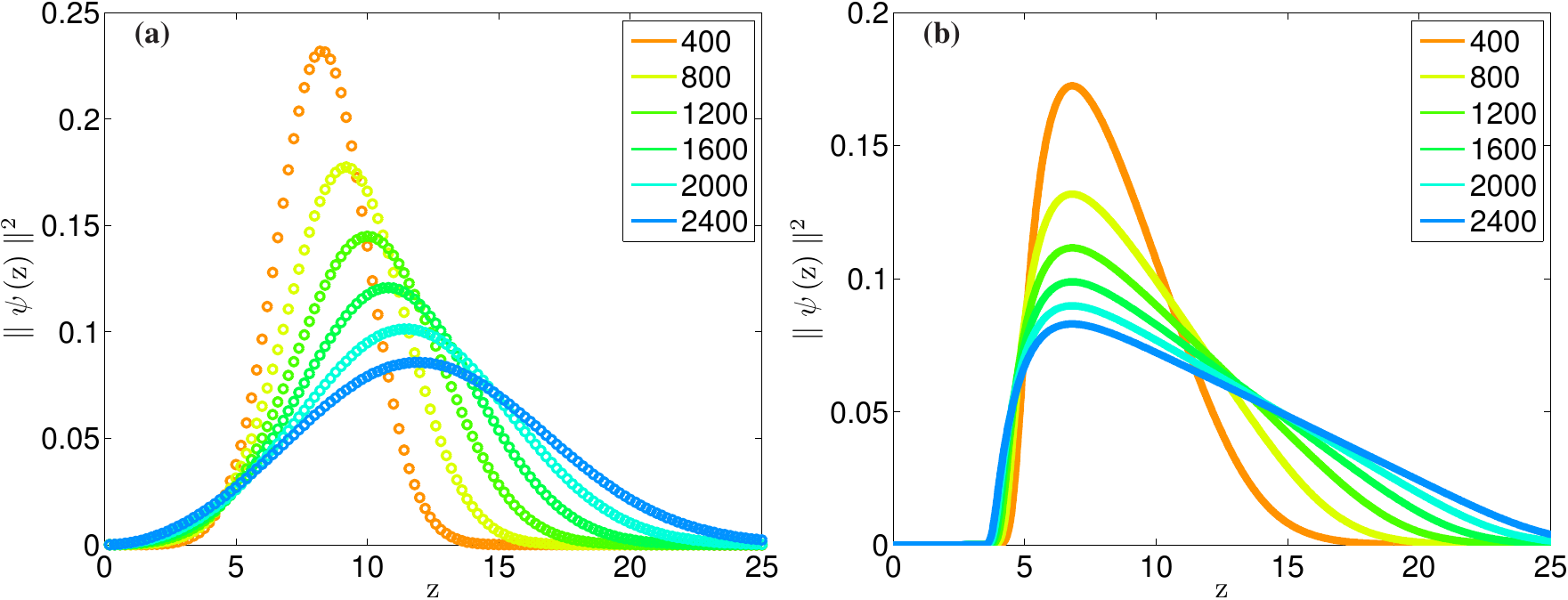} 
\par\end{centering}
\caption{(Color online) Probability density plots from (a) Gaussian ansatz
(\ref{eq:5}) and (b) numerical calculations for an increasing number
of atoms from the top to the bottom.}

\label{LowNum}
 
\end{figure}

\section{Thomas-Fermi Limit Solution}

\label{TF} For a large enough number of atoms, the effective interaction
term and the potential term are much larger than the kinetic term.
In such a case, an approximate Thomas-Fermi (TF) solution is found
by neglecting the much smaller kinetic term. Thus, the time-independent
GPE reduces to 
\begin{equation}
\mu\approx z+V_{0}e^{-z}+\text{G}_{\text{B}}\lvert\psi(z)\rvert^{2}.
\end{equation}
In order to determine the chemical potential, we use the normalization
condition, which reads in the dimensionless scheme as ${\int\left|\psi\right|^{2}dz=1}$.
Thus, we get 
\begin{equation}
1=\frac{\mu}{\text{G}_{\text{B}}}\int_{z_{1}}^{z_{2}}\left(1-\frac{z}{\mu}-\frac{V_{0}}{\mu}e^{-z}\right)dz,\label{normInt}
\end{equation}
where $z_{1}$ and $z_{2}$ denote the zeros of the integrand. For
larger $z$ the decaying exponential vanishes, and so we have $z_{2}\approx\mu$.
Examining the smaller root, we see that for small values of $z$ and
moderate values of $\mu$, ${z/\mu}$ is quite small, thus a reasonable
approximation for this root is ${z_{1}\approx\log\left(V_{0}/\mu\right)}$.
This motivates us to divide the TF solution into two parts: first,
the soft-wall TF solution, where $V_{0}>\mu$, so that $z_{1}$ is
larger than zero. Second, the hard-wall TF solution in the case $V_{0}<\mu$,
where the lower integration limit $z_{1}$ would be less than zero,
so the soft-wall TF wave function would fail. The latter case necessitates
to use the mirror principle in order to guarantee the hard-wall boundary
condition.

\subsection{Soft-wall Thomas-Fermi solution}

With the integration boundaries $z_{1}\approx\log(V_{0}/\mu)$ and
$z_{2}\approx\mu$ known to be a good approximation, we carry out
the integration in Eq. (\ref{normInt}), yielding 
\begin{equation}
2\text{G}_{\text{B}}\approx\mu^{2}\left\{ 1-\frac{2}{\mu}\left[\log\left(\frac{V_{0}}{\mu}\right)+1\right]\right\} ,\label{muEqu}
\end{equation}
where we have neglected the small terms $(V_{0}/\mu)e^{-\mu}$ and
$\log\left(V_{0}/\mu\right)^{2}/(2\mu)$. Thus, to leading order,
we have ${\mu\approx\sqrt{2\text{G}_{\text{B}}}}$ with a subsequent
logarithmic correction. For small changes of the chemical potential,
the natural log term in Eq. (\ref{muEqu}) does not change significantly.
Therefore in Eq. (\ref{muEqu}) we can substitute $\sqrt{2\text{G}_{\text{B}}}$
for $\mu$ in the natural log. Using the quadratic equation and neglecting
small terms in the square root, we obtain the improved approximation
\begin{equation}
\mu\approx\sqrt{2\text{G}_{\text{B}}}+1+\log\left(V_{0}/\sqrt{2\text{G}_{\text{B}}}\right).\label{eq:12}
\end{equation}

Thus, assuming that $V_{0}>\sqrt{2\text{G}_{\text{B}}}$, we obtain
the following soft-wall TF solution for ${z_{1}<z<z_{2}}$. 
\begin{equation}
\psi(z)=\sqrt{\frac{\mu}{\text{G}_{\text{B}}}\left(1-\frac{z}{\mu}-\frac{V_{0}}{\mu}e^{-z}\right)},\label{rightPsi}
\end{equation}
where $\mu$ is given by Eq. (\ref{eq:12}). The function is set to
zero for ${z<z_{1}\approx\log(V_{0}/\sqrt{2G})}$ and ${z>z_{2}\approx\mu}$,
because the probability density cannot be less than zero.

\subsection{Hard-wall Thomas-Fermi solution}

For particle numbers $N_{\text{B}}>2.35\times10^{5}$ the soft-wall
TF solution is not valid anymore as $z_{1}\approx\log(V_{0}/\sqrt{2\text{G}_{\text{B}}})$
becomes negative. In order to extend this approximate solution to
the case where $V_{0}<\mu$, we must work out the corresponding hard-wall
TF solution. With the help of the mirror analogy \cite{Robinett06,Belloni14},
we obtain the approximate TF wave function
 
\begin{equation}
\psi(z)=\left\{ \begin{array}{lr}
\sqrt{\frac{1}{M}}\bigg[\sqrt{\left(\frac{\mu}{\text{G}_{\text{B}}}\right)\left(1-\frac{z}{\mu}-\frac{V_{0}}{\mu}e^{-z}\right)}-\sqrt{\left(\frac{\mu}{\text{G}_{\text{B}}}\right)\left(1+\frac{z}{\mu}-\frac{V_{0}}{\mu}e^{+z}\right)}\hspace{1mm}\bigg] & \text{for }0<z<|z_{1}|\\
\sqrt{\frac{1}{M}\left(\frac{\mu}{\text{G}_{\text{B}}}\right)\left(1-\frac{z}{\mu}-\frac{V_{0}}{\mu}e^{-z}\right)} & \text{for }|z_{1}|<z<z_{2}.
\end{array}\right.
\end{equation}
Here, $M$ denotes the normalization constant which
is determined from $\int\left|\psi\right|^{2}dz=1$. Note that an
analytical derivation of $M$ is not possible, therefore, we performed
the respective integration numerically.

\section{Numerical Methods and Results}

\label{Num} In order to demonstrate the validity of the proposed
analytical results, we numerically find the ground state of the wave
function by propagating the GPE in imaginary time, with the help of
the split operator technique \cite{Javanainen06,Vudragovic12,Kumar15,Loncar15,Sataric16}.
For the above mentioned experimental parameters and with the value
of $N_{\text{B}}$ ranging from $10^{6}$ to $10^{6}$ atoms, the ground-state
energy of the BEC in a GOST does, indeed, quickly converge as shown
in Fig. \ref{converge}.


\begin{figure}[h]
\begin{centering}
\includegraphics[width=5.5in,height=2.3in]{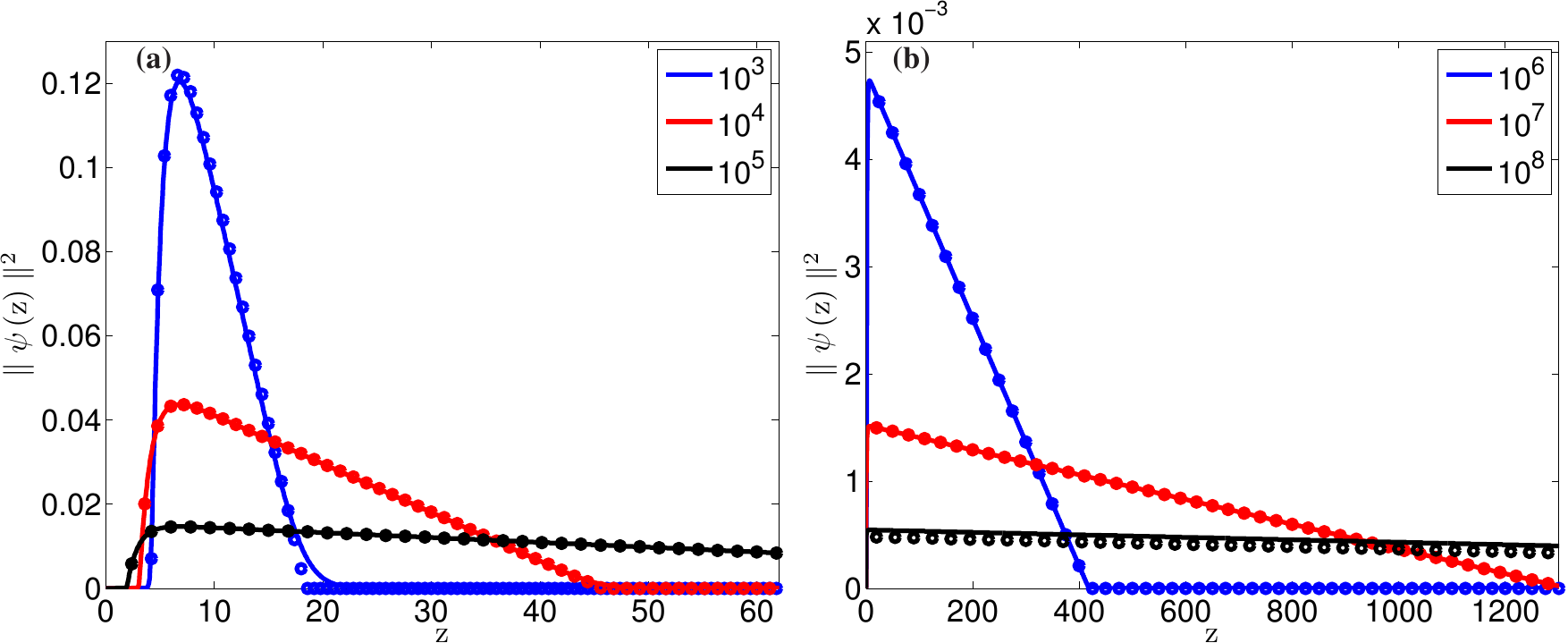} 
\par\end{centering}

\protect\protect\protect\protect\protect\protect\protect\caption{(Color online) Comparison of numerical results from GPE (solid lines)
with TF approximation (circles), (a) $N_{\text{B}}=10^{3}$, $N_{\text{B}}=10^{4}$,
and $N_{\text{B}}=10^{5}$. (b) $N_{\text{B}}=10^{6}$, $N_{\text{B}}=10^{7}$,
and $N_{\text{B}}=10^{8}$ from top to bottom, respectively.}

\label{HighPart1} \vspace{-5mm}
 
\end{figure}

With this technique in mind, we compare our analytical results from
Sec. \ref{Ansatz} and \ref{TF} to numerical results and show how
the BEC wave function in a GOST changes with increasing the number
of atoms. For a smaller number of atoms, the variational Gaussian
ansatz is more suitable as shown in {Fig.~\ref{LowNum}}, whereas
for a larger number of atoms the Thomas-Fermi approximation turns
out to be quite accurate as shown in {Fig.~\ref{HighPart1}}. The
variational Gaussian ansatz roughly reproduces the mean, but it is
rather symmetrical, unlike the numerical results as the number of
atoms approach $2400$, see {Fig.~\ref{LowNum}}. Qualitatively,
the BEC width is proportional to the number of atoms in a GOST. However,
due to the EW decaying exponential potential, the BEC cannot expand
in the negative $z$-direction, so the BEC wave function takes up
a triangular shape for $N_{\text{B}}$ larger than $10^{6}$ as shown
in Fig. \ref{HighPart1}. The agreement between numerical and analytical
TF results is much better for larger number of atoms. Note that the
BEC wave function is quite wide for $10^{8}$ atoms in Fig. \ref{HighPart1}.

\begin{figure}[h]
\begin{centering}
\includegraphics[width=3.4in,height=2.5in,keepaspectratio]{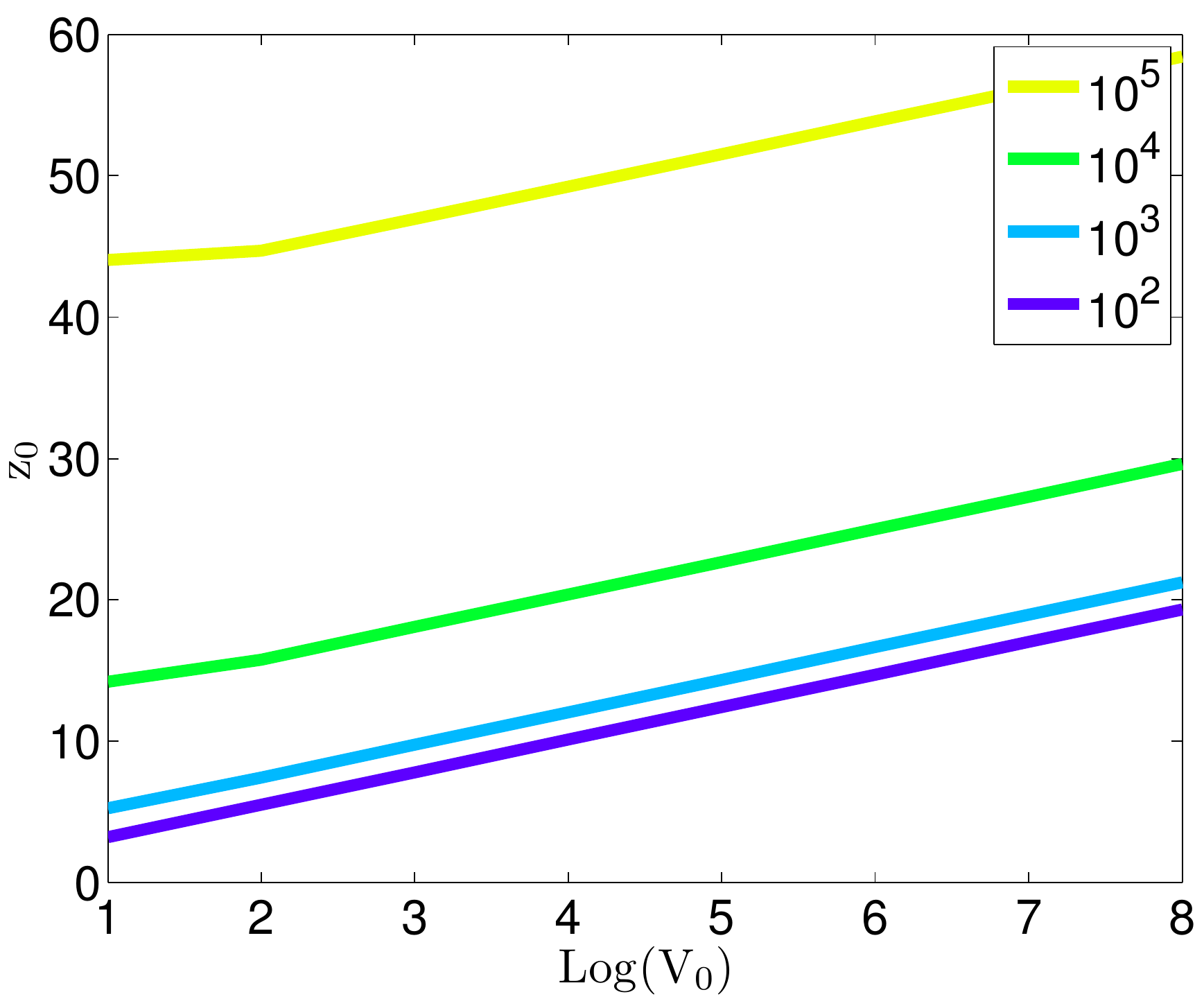} 
\par\end{centering}

\protect\protect\protect\protect\protect\protect\protect\caption{(Color online) Mean position of the BEC versus EW strength $V_{0}$
for decreasing number of atoms from the top to the bottom.}

\label{MeanPos} 
\end{figure}

The van-der-Waal forces with the surface can demolish the BEC, so
it is necessary to have a larger EW potential. Therefore, keeping
in mind recent developments in the laser field technology, it is possible
to increase the EW strength $V_{0}$ by increasing the laser power
\cite{Yin2006}. Thus, we explore now the parameter space of our model.
Irrespective of the number of atoms, the mean position of the BEC
in GOST increases only moderately with $V_{0}$. For the soft-wall
case $V_{0}>\mu$ as shown in Fig. \ref{MeanPos}, the maximum of
the wave function occurs at $\log(V_{0})$, but for the hard-wall
case $V_{0}<\mu$ the maximum of the BEC wave function exists at $\left|\log(V_{0}/\sqrt{2\text{G}_{\text{B}}})\right|$.
Thus, due to the interaction term, for a large number of atoms, the
maximum of the wave function no longer remains within the minimum
of the trapping potential $z_{0}^{{\rm {min}}}\approx\log(V_{0})$.

\begin{figure}[h]
\begin{centering}
\includegraphics[width=3.4in,height=2.5in]{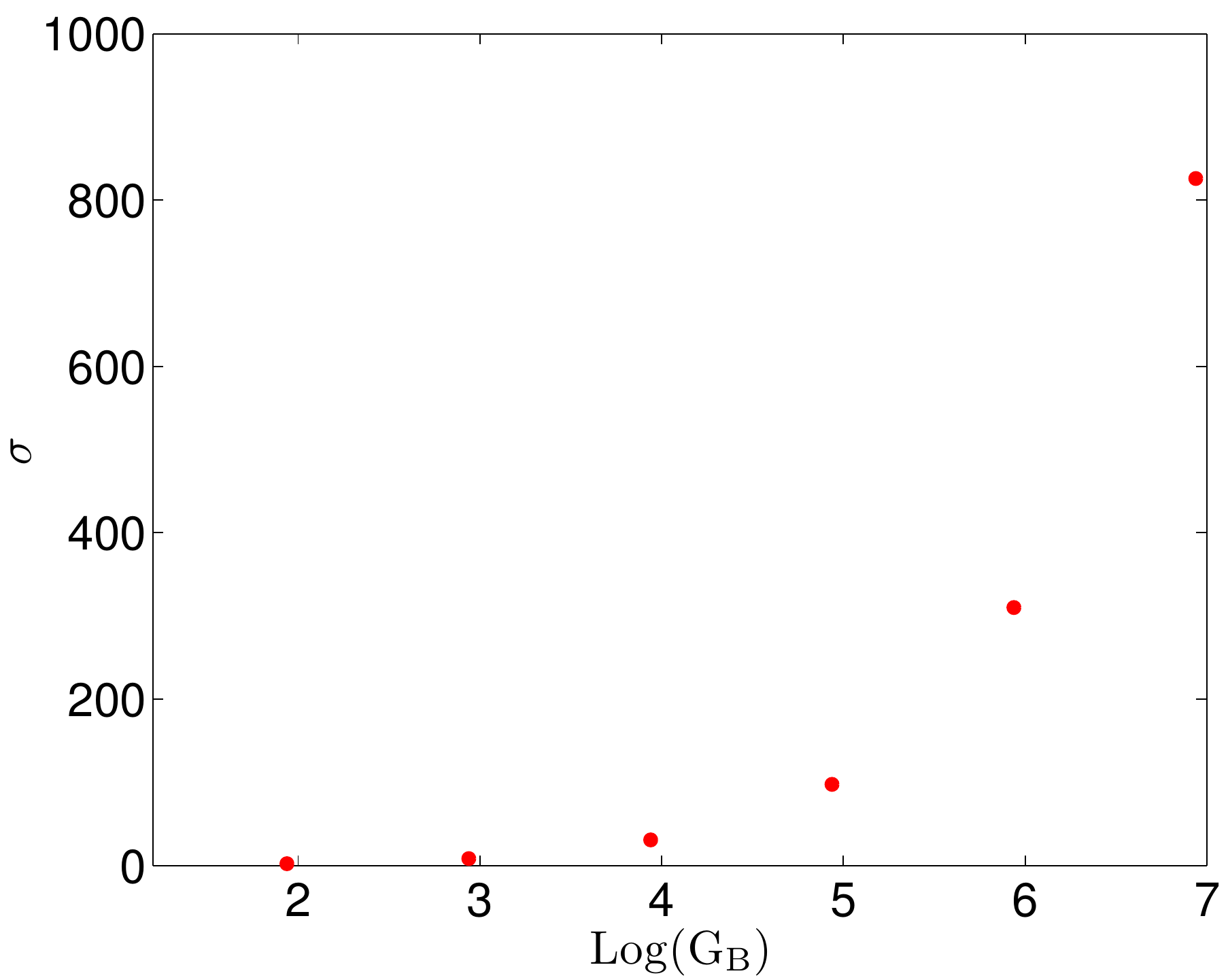} 
\par\end{centering}

\protect\protect\protect\protect\protect\protect\protect\caption{(Color online) Standard deviation $\sigma$ of BEC wave function increases
with inter-particle interaction strength $\text{G}_{\text{B}}$ for
EW strength $V_{0}=906$.}

\label{Width} 
\end{figure}

The BEC wave function in a GOST becomes asymmetric for larger interaction
strengths. Therefore we quantify the BEC width based on the standard
deviation $\sigma=\sqrt{<z^{2}>-<z>^{2}}$, where $<\bullet>=\int\bullet\left|\psi(z)\right|^{2}dz$
denotes the expectation value. As shown in {Fig.~\ref{Width}},
the BEC standard deviation grows extremely rapidly with increasing
number of atoms $N_{\text{B}}$. However, on the other hand, changing
$V_{0}$ only slightly affects the standard deviation $\sigma$.

\section{Time-of-flight (TOF) Expansion}

\label{TOF} The standard observation of a BEC is based on suddenly
turning off the trapping potential and allowing the atoms to expand
non-ballistically. The resulting time-of-flight (TOF) measurements
are performed either by acquiring the absorption signal of the probe
laser beam through the falling and expanding BEC cloud, or by measuring
the fluorescence of the atoms which are excited by a resonant probe
light \cite{Perrin06}.

\begin{figure}[h]
\begin{centering}
\includegraphics[width=3.4in,height=2.5in]{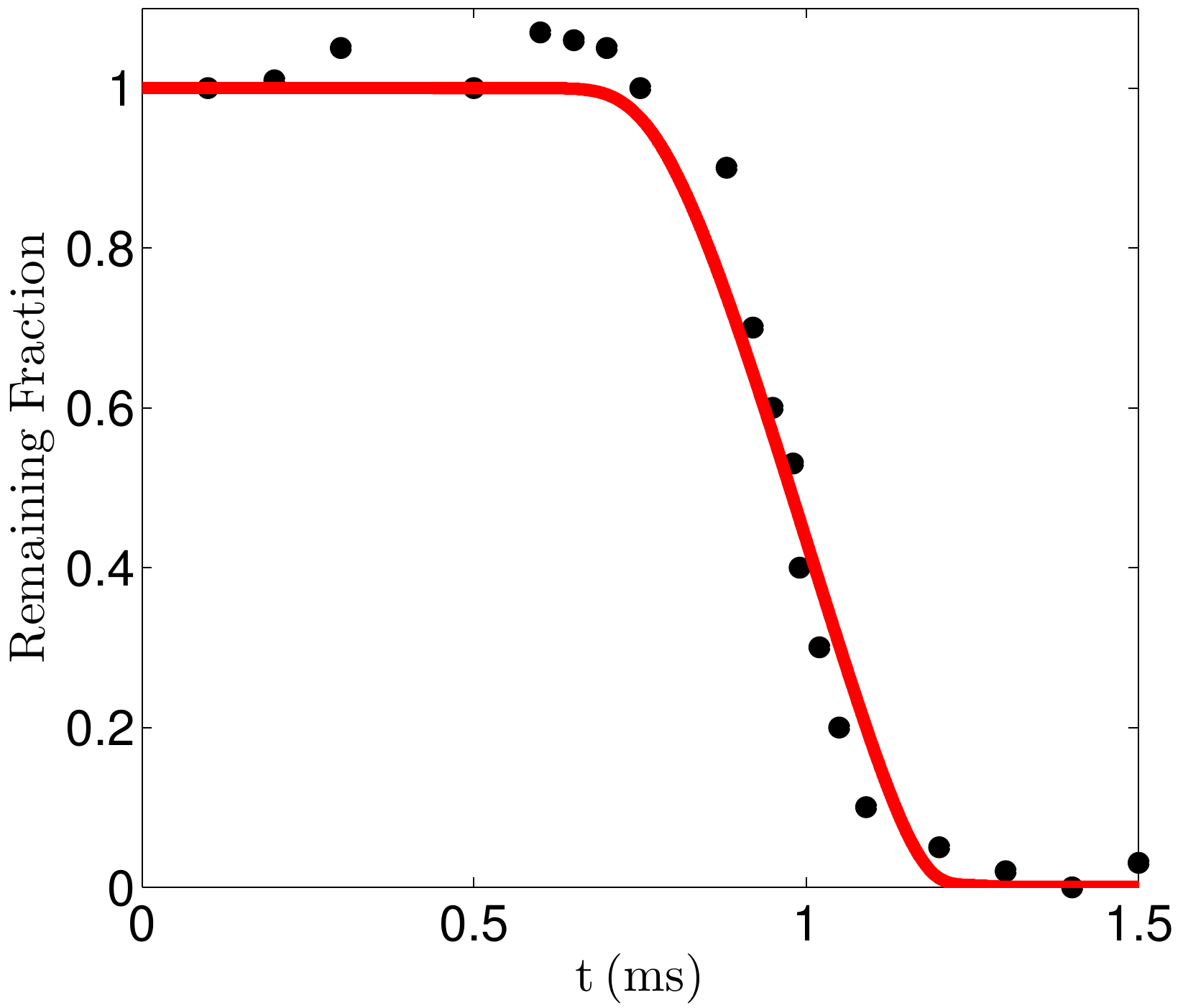} 
\par\end{centering}

\protect\protect\protect\protect\protect\protect\protect\caption{(Color online) Fraction of remaining atoms during time-of-flight during
a vertical expansion for $V_{0}=453$. Initially, total number of
atoms is $N_{\text{B}}=2400$. Here the circles stem from the Innsbruck
experimental \cite{PhysRevLett.92.173003}, whereas the solid line
shows our numerical results.}

\label{reproduced} 
\end{figure}

In the Innsbruck experiment, the remaining number of atoms is measured
after allowing atoms from the GOST to expand vertically by suddenly
turning off the EW \cite{PhysRevLett.92.173003}. Note that some particles
are lost due to thermalization processes which occur when the particles
hit the prism or due to the van-der-Waal forces with the surface.
Although this Innsbruck experiment uses a 2D pancake shaped BEC, when
performing this vertical expansion the transversally confining beam
is kept constant, so our quasi-1D model for BEC should apply in this
case. Using the experimental parameters in Ref. \cite{PhysRevLett.92.173003},
we numerically reproduce their vertical expansion curve (their Figure
2), as shown in {Fig. \ref{reproduced}}. To this end, we use $N_{\text{B}}$
= 2400 atoms and $V_{0}=453$, yielding an initial condensate wave
function with dimensionless standard deviation $\sigma=0.86$ and
the dimensionless mean position $z_{0}=6.40$. We propagate this wave
function without the hard-wall boundary condition. Then we approximate
the fraction of remaining atoms by ${\int_{0}^{\infty}\lvert\psi(z,t)\rvert^{2}dz}$,
as atoms in the BEC wave function extending past $z=0$ are lost by
sticking to the surface. As the interaction term is quite small in
the Innsbruck experiment, the standard deviation of the BEC is small
and remains roughly constant during the time-of-flight, so the loss
of atoms does not affect time-of-flight expansion significantly for
$t<0.7~{\rm {ms}}$ as shown in {Fig. \ref{reproduced}}.

\begin{figure}[h]
\begin{centering}
\includegraphics[width=5.5in,height=2.3in]{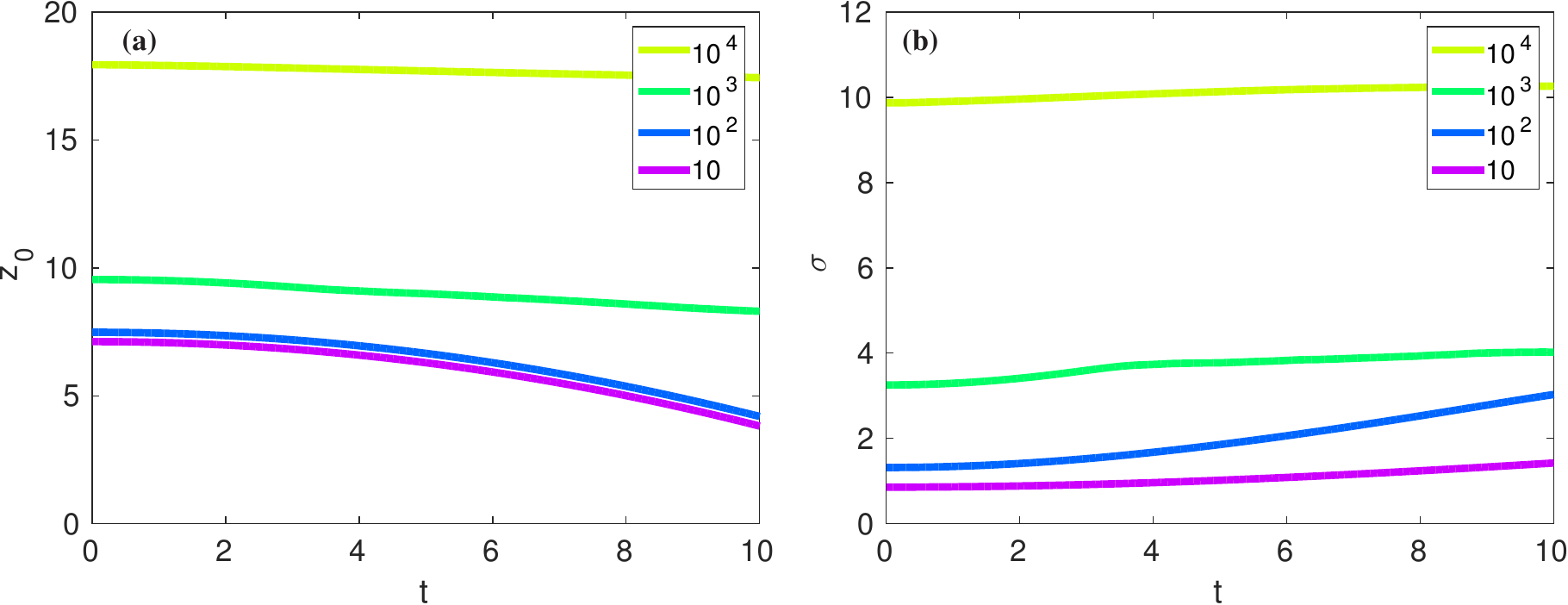} 
\par\end{centering}

\caption{(Color online) (a) Mean position $z_{0}$ and (b) standard deviation
$\sigma$ of BEC density in the time-of-flight expansion for decreasing
number of atoms from the top to the bottom for $V_{0}=906$. }

\label{TOF_Pos} 
\end{figure}

We also simulated the TOF expansion without the prism at $z=0$ for
different particle numbers $N_{\text{B}}$. We see that the mean position
of the BEC drops due to gravity as shown in {Fig.~\ref{TOF_Pos}}
with different rates, which strongly depend upon the number of confined
atoms. At the same time the BEC width, which is proportional to the
standard deviation, increases according to {Fig.~\ref{TOF_Pos}}.

\section{Dynamics of BEC on hard-wall mirror}

\label{Dynamics} 
\begin{figure}[h]
\begin{centering}
\includegraphics[width=4.0in,height=2.5in]{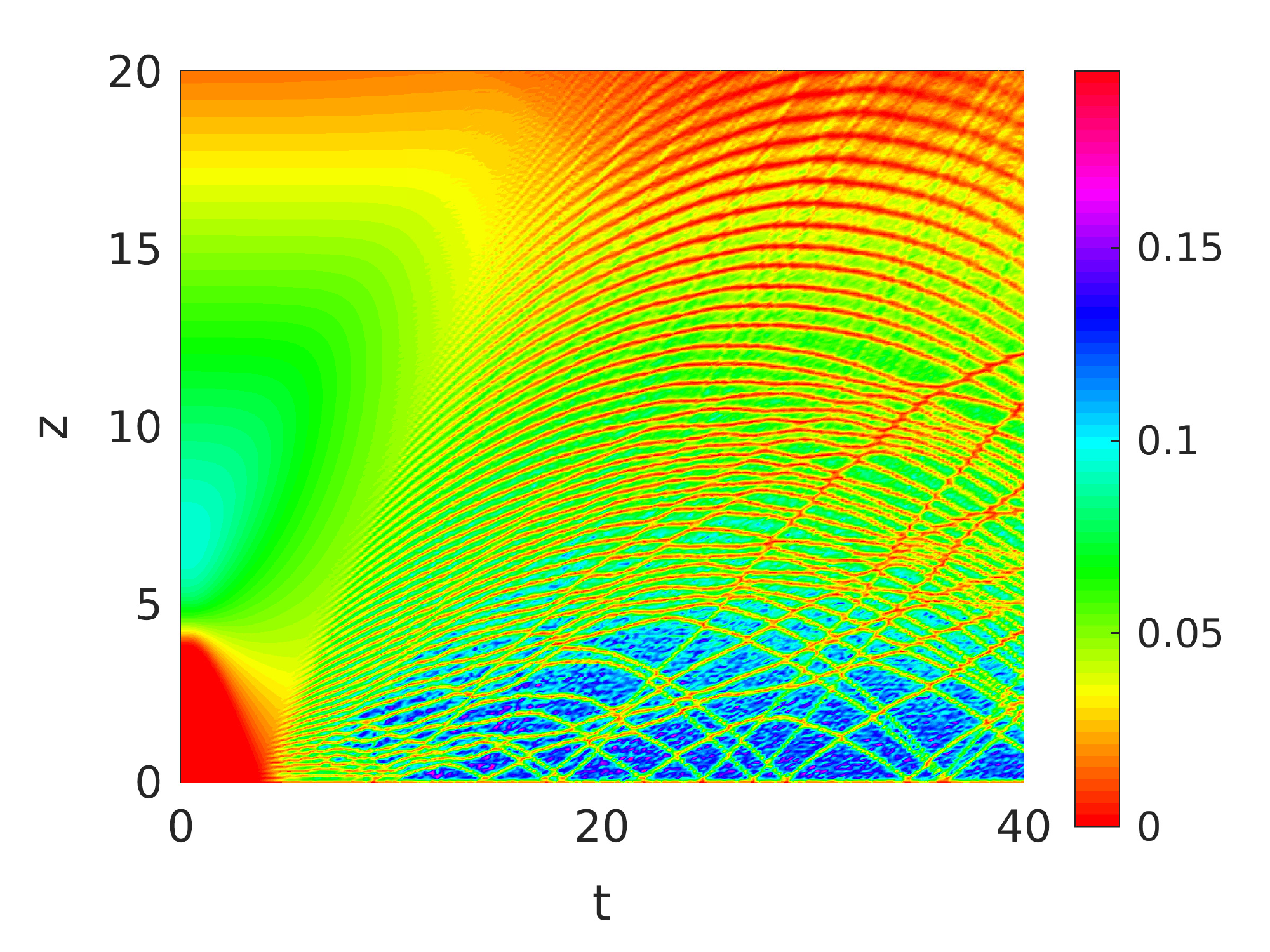} 
\par\end{centering}

\protect\protect\protect\protect\protect\protect\protect\caption{(Color online) Dynamics of BEC in presence of gravity with hard wall
at $z=0$ for $N_{\text{B}}=1800$ number of atoms. Here the color scale represents the density of the BEC. }

\label{Dynamics1} 
\end{figure}

Concerning the dynamics of the BEC in a GOST, we
consider two cases. First we describe the dynamics of the BEC with
$N_{\text{B}}=1800$ after switching off the evanescent potential
and letting the BEC atoms fall on a hard-wall mirror. The latter could
be experimentally realized by a blue-detuned far-off-resonant sheet
of light, and is modeled theoretically via the boundary condition
$V(0)\rightarrow\infty$ at $t>0$. Thus, the BEC of $N_{\text{B}}=1800$
atoms has the dimensionless mean position $z_{0}=10.7$ and the dimensionless
standard deviation $\sigma=4.28$ at time $t=0$. We observe the matter-wave
interference pattern formed upon releasing the condensate from GOST
as shown in Fig. \ref{Dynamics1}, as atoms impinging on the `hard-wall'
at the origin `bounce' back. For short times the atoms remain
near the hard-wall surface, so the BEC dynamics is characterized by
the reflection of atoms from the hard-wall mirror. But for larger
times those atoms, which are far away from the hard-wall mirror, are
reflected above the hard-wall mirror due to collisions among themselves
as shown in Fig. \ref{Dynamics1}. The number of atoms $N_{\text{B}}=1800$
is so large that the BEC is staying quite close to the mirror, so
we have not seen any total reflection of the BEC wave packet.

\begin{figure}[h]
\begin{centering}
\includegraphics[width=3.4in,height=2.4in]{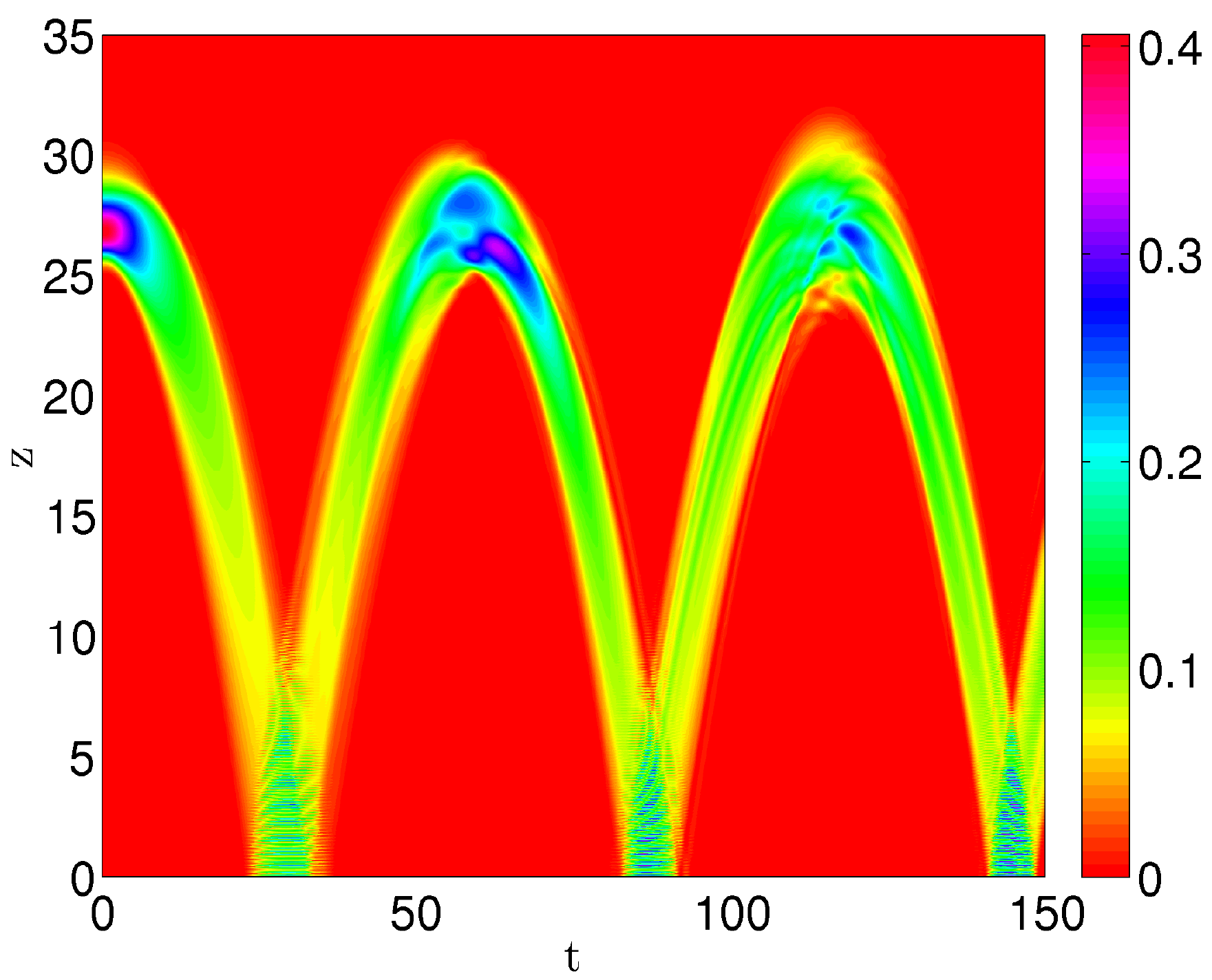} 
\par\end{centering}
\protect\protect\protect\protect\protect\protect\protect\caption{(Color online) Numerical results for the BEC density $\lvert\psi(z,t)\rvert^{2}$
with initial dimensionless mean position $z_{0}=27.2$ and dimensionless
standard deviation $\sigma=0.99$ for $N_{\text{B}}=30$ number of
atoms. BEC experiences full periodic revivals, however incoming and
reflecting BEC wave packets lead to larger matter-wave interference
regions at time intervals t=24-34 and 83-91. Here the color scale represents
the density of the BEC.}
\label{Dynamics2} 
\end{figure}

In order to see the latter scenario, we need a small number of $N_{\text{B}}=30$
particles far away from the hard-wall mirror. This is realized by
the EW potential $V_{0}e^{-\left(z-20\right)}$, which could be implemented
by trapping atoms in a MOT above the surface, so that, once the EW trap is switched off, the atoms have
enough momenta when they hit the hard-wall mirror as shown in Fig. \ref{Dynamics2}. 
Similar to experiments of photonic bouncing balls \cite{PhysRevLett.102.180402}
and plasmonic paddle balls \cite{PhysRevA.84.063805}, the BEC shows
significant self-interference patterns, for example in the time intervals
t=24-34 and 83-91 as shown in Fig. \ref{Dynamics2}, which originate
from the interference of incoming and reflecting BEC wave packets.
It is worth mentioning that a smaller initial width of the BEC wave
packet would lead to finer revivals and a larger initial width of
the BEC would lead to larger interference regions. The evolution of
a BEC falling under gravity and bouncing off a hard-wall mirror formed
by a far-detuned sheet of light was already observed experimentally
by Bongs et al. in both the soft-wall and the hard-wall regime \cite{PhysRevLett.83.3577}.
In the soft-wall regime, they have recorded that the BEC is reflected
up to three times off the optical mirror in the lossy environment.
Due to a large two-particle coupling strength, which in turn results
in a condensate with a larger width, this group also observed a splitting
of the BEC into two parts close to the upper turning point of the
BEC. This effect is heuristically modeled by a GPE dynamics with assuming
that the two-particle interaction strength decreases exponentially
in time. In our case we restricted ourselves to the evolution of the
BEC with a constant coupling constant, so we did not observe any splitting
of the BEC in our simulation, but we do observe the BEC for longer
time intervals as shown in Fig. \ref{Dynamics2}. In our simulation,
we observed more than three reflections of the atomic cloud in a lossless
environment. The center-of-mass of the BEC wave packet shows periodic
revivals with the dimensionless time period $t\approx58.2$ as shown in Fig. \ref{Dynamics2}.
Quantitatively, the classical revival time can be determined in
dimensionless units as $t=2\sqrt{2 z_{0}/k}=56.9$, where $z_{0}$ represents the mean position
of the BEC and the dimensionless energy constant $k$ is defined below
Eq.~(\ref{nondim}).

\section{Conclusion}

In summary, we studied the behavior of a Cs BEC in a quasi-1D optical
surface trap. We have developed approximate solutions to the GP equation
for both small and large numbers of atoms. In the former, we have
used the variational ansatz technique, while in the latter we have
used the Thomas-Fermi approximation. Later on, we compared the analytical
approach with numerical results, which agreed quite well for a wide
range of atom numbers $N_{\text{B}}$.

Furthermore, we have numerically reproduced the experimental result
of Ref. \cite{PhysRevLett.92.173003}, where a 2D BEC is confined
in the radial direction, but is allowed to expand in the vertical
direction freely. This indicates that our analysis could be extended
beyond the 1D case. Our model suggests that for a small particle number
$N_{\text{B}}$ the BEC retains its Gaussian shape in the expansion
and falls due to gravity. As suggested by {Fig~\ref{TOF_Pos}},
for larger number of atoms, the standard deviation does not expand
as fast as compared to small number of particles, therefore we conclude
that the initial number of the particles plays a significant role in
the expansion of the BEC cloud. Afterwards, we investigated the dynamics
of the BEC in the presence of gravity and a hard-wall boundary condition,
where we observed self-interferences and revivals of the wave packet.
The observation of the bouncing of the BEC can be used to characterize
and determine mirror properties such as roughness and steepness. All
our results can be applied to develop atomic interferometers for a
BEC.

\section{Acknowledgment}

We would like to acknowledge Antun Bala\v{z} and Dwight Whitaker for their
insightful comments. We also gratefully acknowledge support from the
German Academic Exchange Service (DAAD). This work was also supported
in part by the German Research Foundation (DFG) via the Collaborative
Research Center SFB/TR49 ``Condensed Matter Systems with Variable
Many-Body Interactions\textquotedblright .

\section*{References}

\bibliographystyle{iopart-num}
\bibliography{1DGOST}

\end{document}